\documentclass[aps,prl,superscriptaddress,preprint,showpacs]{revtex4}
\usepackage{graphics}
\usepackage{epsfig}
\usepackage{latexsym}
\usepackage{colordvi}
\usepackage{amsmath}
\usepackage{amssymb}

\def\lsim{\mathrel{\rlap{\lower3pt\hbox{\hskip1pt$\sim$}}
     \raise1pt\hbox{$<$}}} 
\def\gsim{\mathrel{\rlap{\lower3pt\hbox{\hskip1pt$\sim$}}
     \raise1pt\hbox{$>$}}}

\def\be{\begin{eqnarray}}\def\ba{\begin{eqnarray}}
\def\ee{\end{eqnarray}}\def\ea{\end{eqnarray}}
\def\ben{\begin{enumerate}}\def\bitem{\begin{itemize}}
\def\een{\end{enumerate}}\def\eitem{\end{itemize}}

\begin{document}
\preprint{\parbox[b]{1in}{ \hbox{\tt IC/2008/095 } }}
\title{The Shape of Mesons in Holographic QCD  }

\author{Mahdi Torabian}
\email[E-mail: ]{mahdi@ictp.it} 
 \affiliation{The Abdus Salam International
Centre for Theoretical Physics, Strada Costiera 11, 34014, Trieste,
Italy} \vspace{0.1in}

\author{Ho-Ung Yee}
\email[E-mail: ]{hyee@ictp.it} 
 \affiliation{The Abdus Salam International
Centre for Theoretical Physics, Strada Costiera 11, 34014, Trieste,
Italy} \vspace{0.1in}

\vspace{0.1in}


\begin{abstract}

Based on the expectation that constituent quark model may capture right physics in the large N limit,
we point out that {\it orbital} angular momentum of the quark-antiquark pair inside
light mesons of low spins in the constituent quark model may provide a clue for the holographic dual string model of large N QCD.
Our discussion, relying on a few suggestive assumptions, leads to a necessity of world-sheet fermions in the bulk of dual strings that
can incorporate intrinsic spins of fundamental QCD degrees of freedom.
We also comment on an interesting issue of the size of mesons in the context of holographic QCD.

\end{abstract}

\maketitle
\newpage
\section{}

Since the observation by t'Hooft that the large color $N_c$ limit with $g_{YM}^2 N_c$ finite
should have a dual string theory description \cite{'tHooft:1973jz},
the correct dual string model of large $N_c$ QCD
has evaded much of study done in the last several decades.
Although we still don't have a full understanding of how this theory looks like,
a major progress in addressing this problem has appeared a decade ago
with an idea of holography; AdS/CFT correspondence \cite{Maldacena:1997re}.
Based on the same philosophy, much work has been carried out to
find, or at least to approximate, the dual model of large $N_c$ QCD (see Ref.\cite{Erdmenger:2007cm}
for a review).

In this short note, we point out one observation that may shed some
light on the would-be dual string theory of large $N_c$ QCD.
Our starting point is looking critically at how the QCD mesons of low spins
are described in the current top-down models of holographic QCD
such as the Sakai-Sugimoto model \cite{Witten:1998zw,Sakai:2004cn} (See also Ref.\cite{Kuperstein:2004yk,Casero:2005se}).
In the limit of large color $N_c$, but the number of flavors $N_f$ being kept finite,
one may neglect back-reaction from the quarks to the color dynamics, which is
equivalent to the quenched approximation in lattice QCD.
The corresponding approximation in the holographic QCD models is to treat
the $N_f$ flavor D-branes as probe-branes embedded in the background provided by large $N_c$ color dynamics \cite{Karch:2002sh}.
As the end point of an open string on these flavor branes has a natural interpretation as a dynamical quark
(or an antiquark depending on the orientation of the string),
the open strings living on the world-volume of these probe-branes
can be viewed as composite states of quark-antiquark pair, and hence mesons.
Note that in the dual string description, one only sees color-singlet states, and
these quark-antiquarks are not the bare, color-charged quarks, but rather they are
dressed to be color singlet constituent quarks which carry only flavor charges.
The original bare quarks are expected to be tightly bound as explained in Ref.\cite{Kruczenski:2003be}.
Therefore, one may naturally assign the two end points of a single open string as constituent quark-antiquark pair in the emerging constituent quark model in the large N limit. Our subsequent discussion is based on this hypothetical view-point, which may or may not capture the right qualitative physics for real QCD. This is one potential loop-hole in our following arguments.

Among the open string states on the flavor branes,
of particular interest are the states from the lowest string spectrum
which govern the low energy dynamics of the flavor branes.
Especially, they include a gauge theory with vector fields $A_M$, whose
gauge group corresponds to the global symmetry of QCD such as chiral symmetry.
Eventually, these gauge fields in the holographic QCD give us the pseudo-scalar pions
as well as a tower of spin 1 vector mesons upon the mode expansion to our 4
dimensional spacetime \cite{Kruczenski:2003be,Sakai:2003wu,Babington:2003vm}.
One may also have scalar fields living on the flavor branes which will be responsible for
spin 0 mesons in QCD.
The point is that there is no further higher spin states in the lowest
open string spectrum on the flavor branes.
As they are massless in the holographic 5 dimensions, their 4-dimensional masses
are dictated by the curvature scale of the gravity background upon 4-dimensional reduction.
These masses are parametrically lower than the mass of
states coming from higher excited string spectrum whose scale is given by the effective string scale
in the background.

More specifically, in the case of Sakai-Sugimoto model, the masses of mesons arising from the massless open
string states on the flavor branes behave as $m \sim M_{KK}$ where $M_{KK}$ is the unique scale in the model.
On the other hand, there are other states coming from higher excited level of string modes, which have masses
$m\sim M_{KK}\left(g_{YM}^2 N_c \right)^{1\over 2}$ because their masses are roughly given by the inverse of the
effective string length $l_s^{eff}$ at the IR end point where they are localized, which is given by
$l_s^{eff} \sim {1\over M_{KK}\left(g_{YM}^2 N_c\right)^{1\over 2} }$
and the gravity curvature effects of order $M_{KK}$ to their 4D masses are negligible compared to the above.

The bottom-line is that the light spin 0 and spin 1 mesons, or at least the lightest ones, in holographic QCD
are described by the lowest string modes on the flavor branes.
These include pions $\pi^0, \pi^\pm$, and the lightest spin 1 vector meson, the $\rho$ meson.

There is an interesting implication from this.
In Type IIA string theory that the Sakai-Sugimoto model is based on,
the lowest massless open string states on the flavor branes, that consist of gauge fields and scalars,
arise as
\be
A_\mu \sim \psi_{-{1\over 2}}^\mu |0\rangle_{NS}\quad,\quad \Phi^i \sim
\psi_{-{1\over 2}}^i |0\rangle_{NS}\quad,\label{openstates}
\ee
where $\psi^M$ are the world-sheet fermions in the R-NS formalism. Recall that
in the R-NS quantization scheme we also have world-sheet bosons $X^M(\sigma,\tau)$ which
map the string world-sheet to the (10-dimensional) target space.
Note that it is these
$X^M$ world-sheet fields that describe the spatial shape of open strings in the target space.
We are especially concerned here with the relative displacement $\Delta X^\mu=X^\mu(\pi)-X^\mu(0)$ between
two end points of an open string states at $\sigma=\pi$ and $\sigma=0$,
as they are interpreted as quark-antiquark pair that comprise a meson state.
By looking at the quantum wave-function on $\Delta X^\mu$ of a specific open string state,
we thus should be able to identify the {\it orbital} angular momentum of quark-antiquark pair
inside the corresponding meson state in QCD.

We remind the reader of that our discussion is based on the assumption of constituent quark model in large
N limit, interpreting the two end points as flavor charged constituent quark-antiquark pair.
Therefore, any comparison has
to be made with the known results in the old constituent quark model which has had some qualitative success in
explaining certain quantum numbers of mesons.

As the lowest open string states (\ref{openstates}) in the Sakai-Sugimoto model
do not involve excitations of the modes from $X^\mu$ at all, the corresponding
wave-function on $\Delta X^\mu$ must be spherically symmetric under a rotation in spatial $R^3$.
Therefore, to conform to this specific holographic model for QCD, we conclude that
{\it  light mesons of spin 0 and spin 1 should
have zero {\it orbital} angular momentum $l=0$, } in the framework of constituent quark model,
which can serve as a test of the model.

This is true for a tower of spin 0 and spin 1 mesons
arising from the lowest open string states (\ref{openstates})
on the flavor branes, which would have masses of $m \sim M_{KK}$.
On the other hand, the parametrically heavier mesons with $m\sim M_{KK}\left(g_{YM}^2 N_c \right)^{1\over 2}$
coming from a higher excited string state such as
$\alpha^\mu_{-1}\psi_{-{1\over 2}}^i|0\rangle_{NS}$
would have an orbital angular momentum $l=1$, where $\alpha^\mu_n$ are mode-excitations of $X^\mu$.
This is because {\it orbital} angular momentum is defined to be a quantum number associated
to spatial $R^3$ rotations of $X^a$ ($a=1,2,3$) alone without touching $\psi^\mu$'s, and
it is easily seen that the above states with $\alpha^a_{-n}$ have non-trivial orbital angular momentum.
This implies that
{\it mesons with orbital angular momentum $l\ge 1$ have masses parametrically larger than that of the
$\rho$-meson in strong coupling limit.} In other words, orbital excitations become highly suppressed
in strong coupling.

This conclusion is generic for
any model of holographic QCD based on Type II string theories.
Since the relative orbital angular momentum between
two end points of an open string state is physical, we expect that
our conclusion doesn't depend on the specific quantization scheme of the Type II superstring theories.

It is neither a trivial statement.
Had a model of holographic QCD been a purely bosonic string theory, we would
have had a different conclusion. The massless gauge fields
on flavor branes in a bosonic string theory would be described by $A_\mu \sim \alpha_{-1}^\mu |0\rangle_B$
and the light spin 1 vector mesons out of these states, say $\rho$-meson,
would have orbital angular momentum $l=1$. On the other hand, as the pions are coming from
 $A_z$ where $z$ is the holographic 5'th dimension, we still expect $l=0$ for them.
This is a different characteristic from the models of Type II string theories.
This implies that {\it orbital} angular momentum of quark-antiquark pair
inside mesons, when compared to the constituent quark model, can provide us an important test to corroborate or disprove a given
dual string model of large N QCD.

 When compared to the existing constituent quark model \cite{Glozman:2009rn} and the light-front framework \cite{Brodsky:2006uqa},
the models based on Type II string theories are favored against purely bosonic string models.
The pions and the $\rho$-meson  in these descriptions have zero orbital angular momentum $l=0$.
A dual string model to large N QCD based on a purely bosonic string theory and other models
with similar characteristic seem to contradict to this fact.
To explain spin 1 nature of light vector mesons such as $\rho$-meson, we need additional degrees
of freedom in addition to $X^\mu$ that can carry angular momentum. In the case of Type II string models,
{\it world-sheet fermions} provide this additional angular momentum contribution.
We point out that we are not advocating any type of target-space supersymmetry, as QCD does not
have any element of supersymmetry. However, our observation indicates
a need for {\it world-sheet} fermions that can contribute to angular momentum, with a necessary
{\it world-sheet} superconformal symmetry
 for a consistent string theory with world-sheet fermions. This seems in line with a recent
approach by Thorn for the dual string theory of large N QCD \cite{Thorn:2008ay}.

The {\it total} angular momentum, or total spin, of a meson would be a sum of the orbital angular momentum
of quark-antiquark pair and other contributions such as intrinsic spin of individual quark or antiquark
and spins of gluons. In the dual string theory point of view, these other contributions
to the total spin of an open string state
should be partially attributed to $\psi^\mu$ modes as in the case of massless open strings on flavor branes
in Type II string theories. We will give a plausible speculation on how
these additional sources to the spin of mesons in QCD may
be described by fermionic world-sheet degrees of freedom
such as $\psi^\mu$ in a dual string theory.

Mesons in a dual string model are open string states whose two end points we identify as quark-antiquark pair,
while the bulk of string world-sheet must be composed of gluonic degrees of freedom.
Although gluons as well as quark-antiquark pair would generically have {\it orbital} angular momentum which
should be eventually described by $X^\mu(\sigma,\tau)$ world-sheet fields in the dual string model,
they also possess their intrinsic spins as well; spin 1 for gluons and spin $1\over 2$ for quark and antiquark.
To represent these additional spin degrees of freedom,
one may imagine a toy model of spin 1 chain along the bulk of open string world-sheet
with additional spin $1\over 2$ degrees of freedom on the two boundaries.
The system looks similar to a string-bit type model for weakly coupled gauge theory,
and as we take a strong coupling limit we would expect a continuum limit of this spin chain
to be a more relevant description.

The continuum limit of spin 1 chain has been studied by Inami-Odake long time ago \cite{Inami:1992ap},
which turns out to be the supersymmetric sine-Gordon theory.
It is a deformation of $su(2)$ Wess-Zumino-Witten model of level 2 with central charge $c={3\over 2}$,
which can be realized by a triplet of world-sheet fermions $\psi^a$ ($a=1,2,3$) \cite{zam}.
This is tantalizingly similar to our world-sheet fermions $\psi^\mu$ which presumably
appear in a correct relativistic treatment of the problem.
We also conjecture that the additional spin $1\over 2$ degrees of freedom at the two boundaries
effectively force us to take only NS-type boundary condition in the continuum limit, which
will explain the absence of fermionic mesons in the real QCD.

{\it  Issue of operational definition of orbital angular momentum}

 As the orbital angular momentum is not a symmetry-related conserved observable, there is an issue of how we define it operationally, especially in the field theory side. Regarding this,  an interesting observation made
by ref.\cite{Glozman:2009rn} is that in the theory of relativistic quarks in the constituent quark
approximation, there are two "field theory" operators that can describe vector states at the rest frame (so our
focus is on the $\rho$-mesons), $\bar q \gamma^i q$ and $\bar q\sigma^{0i} q$. It can be shown that each
operators create quark-antiquark pair in some mixture of orbital angular momenta, or more precisely in $^{2S+1}
L_J$-notation (see the equation (2) in ref.\cite{Glozman:2009rn})
\begin{eqnarray}
\bar q  \gamma^i q = \sqrt{2\over 3} | ^3 S_1\rangle +\sqrt{1\over 3} | ^3 D_1\rangle \quad,\quad
\bar q  \sigma^{0i} q =\sqrt{1\over 3} | ^3 S_1\rangle -\sqrt{2\over 3} | ^3 D_1\rangle\quad,
\end{eqnarray}
Although the above relation has to be renormalized to be defined meaningfully, we can use this as the
operational definition of orbital angular momentum at a given scale as the ref.\cite{Glozman:2009rn} has done.
More precisely, the state $ | ^3 S_1\rangle $ "at a scale $\mu$" is defined by the state created by the operator
$\sqrt{2\over 3}\bar q  \gamma^i q +\sqrt{1\over 3}\bar q  \sigma^{0i} q$
which is renormalized at the scale $\mu$. Vice versa for $ | ^3 D_1\rangle $. This provides one useful
operational definition of orbital angular momentum in field theory language, which has in fact been used in
ref.\cite{Glozman:2009rn}. The lattice QCD results in ref.\cite{Glozman:2009rn} is that at low energy IR, the
$\rho$-meson state is dominantly the $ | ^3 S_1\rangle $ state defined as in the above.

What does this imply in regard to our observation in the holographic QCD?
The above scale dependence is in fact easily encoded in holographic model as a dependence
on the "energy direction" or the 5'th radial holographic direction corresponding to the energy scale.
At a given scale or equivalently at a given 5'th coordinate point, we have two choices of making a vector state;
\begin{equation}
\alpha_{-1}^\mu |0\rangle =|D\rangle\quad,\quad \psi_{-{1\over 2}}^\mu |0\rangle = |S\rangle\quad,
\end{equation}
Note that since we don't know the precise dual string theory of large N QCD, we are simply indicating that we can either excite bosonic field or fermionic field to have different orbital wavefunctions of two end-points of the open string of total spin $J=1$.
The above field theory discussion then tells us that, the bulk mode, we tentatively call $A_\mu$, that describes
the $\rho$-meson should be in general a linear combination of $\alpha_{-1}^\mu |0\rangle$ and $\psi_{-{1\over 2}}^\mu |0\rangle$ whose coefficients is {\it dependent} on the 5'th position or the energy scale. It seems that
this should be what really happens in the yet-unknown true holographic dual theory of large N QCD. What do we
have in current Type II based models? There we have only $\psi_{-{1\over 2}}^\mu |0\rangle$ at "all scales", that
is, $\rho$-mesons are dominantly $S$-wave at all scales. This is understandable in regard to the results in
ref.\cite{Glozman:2009rn} because at IR QCD is strongly coupled and current holographic models are also
describing strongly coupled field theory at all scales ( like in D3/D7 models), so that IR property in
ref.\cite{Glozman:2009rn} seems consistent with our observation of $S$-wave of $\rho$-mesons from current
holographic models. This also implies that bosonic string theory cannot be a proper dual string theory of large
N QCD as it can not accommodate S-wave property of $\rho$-meson at IR strong coupling.

Another important question is how practically one can measure the orbital angular momentum in experiments. At
present, it seems to us it is hard to give a model-independent way of probing orbital angular momentum. However,
within a model such as current holographic QCD models, it can be useful dynamical concept that distinguish
different predictions on certain physical observables, like magnetic dipole moment. For example, $l=1$ state
would give additional contribution to the magnetic dipole moment from orbital motion, than the spherically
symmetric $l=0$ state. However, we caution that the quantitative predictions usually depend on the model
details, and in that sense one should take the orbital angular momentum more carefully. 

{\it Comments on the size of mesons }

Not only the shape but also the size of mesons would be an interesting physical
quantity that can be probed experimentally through form factors. As the mesons
are described by open string states on the flavor branes in the models of holographic
QCD, one might hope to
calculate their size from the known string states such as (\ref{openstates}).
We will be concerned only about mesons with low spins, including stringy excited states,
whose appropriate descriptions are in terms of
low lying string spectrum on the flavor branes, while
the relevant study of the mesons in the Regge trajectory with sufficiently high spins
and masses would require analysis via semi-classical strings attached on the flavor branes \cite{Kruczenski:2003be}.
 We emphasize that the size of quark-antiquark pair that we will discuss are that of
color-singlet constituent quarks  in the constituent quark model in large N limit, and one expects not to see original tightly-bound bare quarks
in the dual string description.

A naive expectation on the size of the low-lying string states would be the effective string scale $l_s^{eff}$
measured at the position where the string states are localized. As an example, the string-frame metric of the
Sakai-Sugimoto model takes a form $ds_{10}^2 = e^{2A(U)}\left(dx^\mu\right)^2 +\cdots$ where $U\ge U_{KK}$ is the holographic dimension corresponding to the energy scale
with $U=U_{KK}$ being the IR end point, and $x^\mu$ is the (3+1)D Minkowski coordinate of the QCD field theory.
We stress that it is the size measured with respect to $x^\mu$ that has a correct field theory interpretation.
Therefore, a string state with a rough size $l_s$ measured by the above string-frame metric would
have a size $l_s^{eff}$ in the QCD field theory side as $l_s^{eff}=e^{-A(U)} \cdot l_s$ which
depends on the position $U$ where the string is localized. Through the same vine, a state with a characteristic
10-dimensional energy $E$ would have an effective 4-dimensional energy $E^{eff}=e^{A(U)}\cdot E$ at a position
$U$. As the warp factor $e^{A(U)}$ takes its minimum at $U=U_{KK}$, $e^{A(U_{KK})} \sim {M_{KK}l_s\left(g_{YM}^2
N_c\right)^{1\over 2} }$, every string state tries to localize at the IR tip $U=U_{KK}$ to minimize its energy,
which gives us a rough estimate of its size as \be \Delta X \sim l_s^{eff}(U_{KK}) \sim {1\over
M_{KK}}\left(g_{YM}^2 N_c\right)^{-{1\over 2}}\quad. \label{size} \ee

Although this seems reasonable for the mesons
corresponding to massive excited open string spectrum on the flavor branes
in view of their mass behavior
$m\sim M_{KK}\left(g_{YM}^2 N_c \right)^{1\over 2}\sim {1\over \left(\Delta X\right)}$
one might expect a different answer for the mesons arising from
the lowest string spectrum (\ref{openstates}) whose characteristic
masses are of order of $M_{KK}$.
Indeed, these states have wave-function spread over $U\ge U_{KK}$
and the calculations of their physical form factors such as electromagnetic form factor
give us the effective size of $M_{KK}^{-1}$ \cite{Sakai:2005yt}.
Similar phenomena happen in the case of baryons \cite{Hong:2007kx}. Therefore, our estimate on the size
applies only to the stringy massive states, whose masses are dominated by their string mass, not by the
curvature scale of the background geometry. It would be interesting to be able to compute form factors for these
states to check our claim.

The expression (\ref{size}) looks agreeable as it implies that the size of hadrons becomes smaller when the
coupling gets stronger and more binding. However, this intuition breaks down in the case of $D3/D7$ system in
strong coupling whose low-energy dynamics is a ${\cal N}=2$ deformation of ${\cal N}=4$ $SU(N_c)$ SYM theory
with additional fundamental hypermultiplets mimicking quarks \cite{Karch:2002sh}. Its dual gravity background
for large $N_c$ limit looks as $ds_{10}^2 = {r^2 \over \left(g_{YM}^2 N_c\right)^{1\over 2} l_s^2
}\left(dx^\mu\right)^2 +\cdots$ where $r$ is the holographic dimension corresponding to the energy
scale by $r= l_s^2 E$. Especially the mass of the fundamental hypermultiplets arising from $3-7$ strings is
given by $m_q={\left(\Delta r\right)\over l_s^2 }$ where $\left(\Delta r \right)$ is a distance between $D3$ and
$D7$ branes. The open strings on the flavor $D7$ branes are naturally mapped to the mesonic states of
quark-antiquark bound states in the field theory side, and these open string states are localized around the
position $r=\left(\Delta r\right)$ to minimize their energy. The resulting size estimate of mesons is then \be
\Delta X \sim {\left(g_{YM}^2 N_c\right)^{1\over 4} l_s \over \left(\Delta r\right)}\cdot l_s \sim
{\left(g_{YM}^2 N_c\right)^{1\over 4}\over m_q}\quad, \ee which {\it grows} as the coupling becomes stronger.
Recall that this strange behavior of size is for the color-singlet constituent quarks. This is a puzzling
question for future works. We also comment that even for the lowest mesons coming from the 5D massless
states, the reliable form factor computations such as in ref.\cite{Hong:2003jm} gives us the size
${\left(g_{YM}^2 N_c\right)^{1\over 2}\over m_q}$ which also grows in the strong coupling.

Our final comment is that a {\it quantitative}
calculation of the size of mesons from quantized string states such as (\ref{openstates})
is not under control of our current understanding of string theory.
This is because
a naive calculation of
the size of strings in {\it flat space} seems logarithmically divergent \cite{Karliner:1988hd}.
Indeed this was one of the historic reasons why the string theory was abandoned as
a theory of large N QCD. This question however takes a new twist in the
gauge/gravity correspondence as it advocates dual string theory
description for large N field theories, and the size of hadrons {\it has to} be
finite \cite{Polchinski:2001ju}. Unless we come to know how to quantize string theory
in a curved background such as $AdS_5 \times S^5$,
the precise calculations of hadron size will remain as an open problem.

\subsection*{Acknowledgments}
We would like to thank Kumar S. Narain for
many valuable discussions that lead us to this work.
We also thank M.H. Sarmadi for helpful comments.



\begin{thebibliography}{99}

\bibitem{'tHooft:1973jz}
  G.~'t Hooft,
  Nucl.\ Phys.\  B {\bf 72}, 461 (1974).


\bibitem{Maldacena:1997re}
  J.~M.~Maldacena,
  Adv.\ Theor.\ Math.\ Phys.\  {\bf 2}, 231 (1998).


\bibitem{Erdmenger:2007cm}
 J.~Erdmenger, N.~Evans, I.~Kirsch and E.~Threlfall,
  Eur.\ Phys.\ J.\  A {\bf 35}, 81 (2008).

\bibitem{Witten:1998zw}
  E.~Witten,
  Adv.\ Theor.\ Math.\ Phys.\  {\bf 2}, 505 (1998).


\bibitem{Sakai:2004cn}
  T.~Sakai and S.~Sugimoto,
  Prog.\ Theor.\ Phys.\  {\bf 113}, 843 (2005).

\bibitem{Kuperstein:2004yk}
  S.~Kuperstein and J.~Sonnenschein,
  JHEP {\bf 0407}, 049 (2004).

\bibitem{Casero:2005se}
  R.~Casero, A.~Paredes and J.~Sonnenschein,
  JHEP {\bf 0601}, 127 (2006).



\bibitem{Karch:2002sh}
  A.~Karch and E.~Katz,
  JHEP {\bf 0206}, 043 (2002).

\bibitem{Kruczenski:2003be}
  M.~Kruczenski, D.~Mateos, R.~C.~Myers and D.~J.~Winters,
  JHEP {\bf 0307}, 049 (2003).

\bibitem{Sakai:2003wu}
  T.~Sakai and J.~Sonnenschein,
  JHEP {\bf 0309}, 047 (2003).

\bibitem{Babington:2003vm}
  J.~Babington, J.~Erdmenger, N.~J.~Evans, Z.~Guralnik and I.~Kirsch,
  Phys.\ Rev.\  D {\bf 69}, 066007 (2004).

\bibitem{Thorn:2008ay}
  C.~B.~Thorn,
  arXiv:0808.0458 [hep-th].


\bibitem{Inami:1992ap}
  T.~Inami and S.~Odake,
  Phys.\ Rev.\ Lett.\  {\bf 70}, 2016 (1993).

\bibitem{zam}
A.~Zamolodchikov and V.~A.~Fateev, Sov. J. Nucl. Phys. 32 (1981) 298.


\bibitem{Glozman:2009rn}
  L.~Y.~Glozman, C.~B.~Lang and M.~Limmer,
  arXiv:0905.0811 [hep-lat].

\bibitem{Brodsky:2006uqa}
  S.~J.~Brodsky and G.~F.~de Teramond,
  Phys.\ Rev.\ Lett.\  {\bf 96}, 201601 (2006).


\bibitem{Sakai:2005yt}
  T.~Sakai and S.~Sugimoto,
  Prog.\ Theor.\ Phys.\  {\bf 114}, 1083 (2005).

\bibitem{Hong:2007kx}
  D.~K.~Hong, M.~Rho, H.~U.~Yee and P.~Yi,
  Phys.\ Rev.\  D {\bf 76}, 061901 (2007);
  H.~Hata, T.~Sakai, S.~Sugimoto and S.~Yamato,
  arXiv:hep-th/0701280;
  D.~K.~Hong, M.~Rho, H.~U.~Yee and P.~Yi,
  JHEP {\bf 0709}, 063 (2007).

\bibitem{Hong:2003jm}
  S.~Hong, S.~Yoon and M.~J.~Strassler,
  JHEP {\bf 0404}, 046 (2004).


\bibitem{Karliner:1988hd}
  M.~Karliner, I.~R.~Klebanov and L.~Susskind,
  Int.\ J.\ Mod.\ Phys.\  A {\bf 3}, 1981 (1988).

\bibitem{Polchinski:2001ju}
  J.~Polchinski and L.~Susskind,
  arXiv:hep-th/0112204.



\end{thebibliography}
\end{document}